# Intrinsic degradation mechanism of nearly lattice-matched InAlN layers grown on GaN substrates

Guillaume Perillat-Merceroz, Gatien Cosendey, Jean-François Carlin, Raphaël Butté, and Nicolas Grandjean

*Institute of Condensed MatterPhysics (ICMP), École Polytechnique Fédérale de Lausanne (EPFL), CH-1015 Lausanne, Switzerland*

Thanks toits high refractive index contrast, band gap and polarization mismatch compared to GaN, $In_{0.17}Al_{0.83}N$ layers lattice-matched to GaN are an attractive solution for applications such as distributed Bragg reflectors, ultraviolet light-emitting diodes, orhigh electron mobility transistors. In order to study the structural degradation mechanism of InAlN layers with increasing thickness, we performed metalorganic vapor phase epitaxy of InAlN layers of thicknesses ranging from 2 to 500 nm, on free-standing (0001) GaN substrates with a low density of threading dislocations, for In compositions of 13.5% (layers under tensile strain), and 19.7%(layers under compressive strain). In both cases, a surface morphology with hillocks isinitially observed, followed by the appearance of V-defects. We propose that those hillocks arise due to kinetic roughening, and that V-defects subsequently appear beyonda critical hillock size. It is seen that the critical thickness for the appearance of V-defects increases together with the surface diffusion length either by increasing the temperature or the In flux because of asurfactant effect.In thick InAlN layers, a better (worse) In incorporation occurring on the concave (convex) shape surfaces of the V-defects is observed leading to a top phase-separated InAlN layer lying on the initial homogeneous InAlN layerafterV-defects coalescence.It is suggested that similar mechanisms could be responsible for the degradation of thick InGaN layers.

## I. INTRODUCTION

The InAlN alloy was not very much studied compared to other III-nitrideternary alloys such as InGaN and AlGaN, up to the demonstration of high reflectivity crack-free nearly lattice-matched InAlN/GaN distributed Bragg reflectors (DBRs),[1] andGaN/InAlN high electron mobility transistors.[2–5]It was mainly due to the difficulty of growing homogeneous InAlN layers, because of the very different bond length and growth temperature of the InN and AlN binary compounds. Other applications have been proposed since then, taking advantage ofthe large refractive index contrast with respect to GaN, the large polarization mismatch, the large bandgap (4.5 eV), and the possibility to grow $In_{0.17}Al_{0.83}N$ layers lattice matched to GaN, *i.e.* with no defects due to strain relaxation. GaN/InAlN multiple quantum well structures for near-infrared intersubband applicationshavebeen reported.[6]Moreover, InAlN has been used for the realization of ultraviolet photodiodes,[7] and cladding layers in edge emitting laser diodes.[8,9]Recently, *p*-type doping of InAlN layers was demonstrated.[10]Finally, an optically-pumped verticalexternalcavity surface emitting laser,[11] and anelectrically-pumped monolithic vertical cavity surface emitting laser[12] have been demonstrated with the use of a bottom InAlN/GaN DBR grown on free-standing (FS) GaN substrate. All these applications are reviewed in detail in Refs.13 and 14.

Because the structural properties of InAlN layers have a strong impact on electrical and optical ones, and thus on the device efficiency, these have been more and more studied lately.Metalorganic vapor phase epitaxy (MOVPE) is the commonly used technique to grow InAlN layers and make devices. Generally, a GaN template is grown on a foreign substrate, like sapphire, silicon or SiC.The nearly lattice-matchedInAlN layers grown on top ofsuch a template usually contain defects like hillocks whoseorigin is unclear,[15–19] dislocations natively present in the GaN template because of GaN growth performed on a foreign substrate, and V-defects. V-defects (or V-pits) are inverted empty pyramids with a hexagonal base, and they aregenerally attributed to threading dislocations.[20–23] In addition to the above-mentioned structural defects, InAlN layers were shown to degrade with increasing thickness.[24]Thus homogeneous layers with a given composition give birth to an upper layer with a higher or a lower indium composition.[25–30]Some authors have proposed a mechanism for this degradation occurring with increasing thickness.[29,30]It was suggested that threading dislocations due to the heteroepitaxial growth of the GaN template on sapphire were the only cause for the formation of V-defects. It was then proposed that the coalescence of V-defects led to the growth of an upper InAlN layer with less indium than the original bottom InAlN layer, because of poorer indium incorporation occurring on the inclined facets of the V-defects.

In the present study, we aim at providing a better understanding of theformation mechanismof structural defects, and of the overall degradation mechanism of thick InAlN layers. For this purpose,the growth of $In_xAl_{1-x}N$ layerswas directly performed on FS GaN substrates, which contain a low density of threading dislocations, in order to avoid structural degradation which is thought to be due to the heteroepitaxial growth of GaN on a foreign substrate. Structural defects (hillocks and V-defects) were observed indifferently for In compositions of 13.5% (*i.e.*, for layers under tensile strain), and of 19.7% (*i.e.*, forlayersunder compressive strain), suggesting that the origin of these defects is not related to strain.Contrary to the common belief, it is shown that V-defects are present even if threading dislocations are absent.Moreover, phase separation occurs because of the growth on these V-defects:concave parts of the V-defects are shown to be In-rich, while convex parts are In-poor. When V-defects coalesce, the upper layer shows fourphases: In-rich and In-poor walls, InAlN with the nominal composition, and columnar voids. The degradation mechanism of thick InAlN layers is thus intrinsic and is neither due to theheteroepitaxy on sapphire nor to strain. These observations of the degradation of thick nearly lattice-matched InAlN layers grown on high quality FS GaN substrates could also help understanding the degradation of thick InGaN layers, which is presently attributed to strain relaxation.[31]

## II. EXPERIMENTAL DETAILS

MOVPEgrowth of $In_xAl_{1-x}N$ layers was performed on FS GaN substrates, for In compositions of 13.5% and19.7%. The lattice-matched composition being achievedfor an In content close to 17%,[32] the 13.5% layers are under tensile strain, whereas the 19.7% ones are under compressive strain. The optimal growth temperature of InN layers grown by MOVPE is around 600°C, whereas it is around 1100°C for AlN. At the growth temperatures considered here (840°C for an indium content of 17%), it is difficult to incorporate In, and the surface diffusion length of adatoms is short, especially for Al. The indium composition was mainly tunedby varying the growth temperature: for a 5°C increase in the temperature, the indium content decreases by 1%. The In/Al flux ratio also has an impact on the indium composition, and it was shown to be an important parameteraffecting the structural properties.[33] This ratio was fixed to 3.5 in almost all the samples we studied here, which gave the best results in term of structural quality. One sample with a ratio of 1.8 is also shown for comparison.The FS GaN substrateshave a nominal dislocation density less than$4\times10^7$ cm$^{-2}$.The growth rate was calibrated to estimate the thicknesses of the layers according to the growth time. Measured thicknesses by transmission electron microscopy (TEM) can be slightly different.The indium compositionsweredetermined by high-resolution x-ray diffraction. Information on the surface morphology of the layers was obtained from atomic force microscopy (AFM) measurements. TEM was performed for cross-sections and plan-viewson a FEI Osiris microscope operated at 200 kV. Scanning TEM (STEM) images taken with a high-angle annular dark field detector mainly exhibit a chemical contrast: a brighter contrast means that the region of interest is either thicker or contains a heavier element. They are usually called Z-contrast images, Z being the atomic number of a chemical element. Energy dispersive x-ray spectroscopy (EDX) mapping was also performed on this instrument. Sample preparation for TEM measurements was done by the wedge polishing technique: samples were thinned by mechanical and mechanico-chemical polishing with a wedge shape. In such a way, the defects potentially introduced by ion milling (classically used for TEM lamella thinning) are avoided.

## III. RESULTS
### A. Observation of hillocks and V-defects
#### 1. Role of the InAlN layer thickness

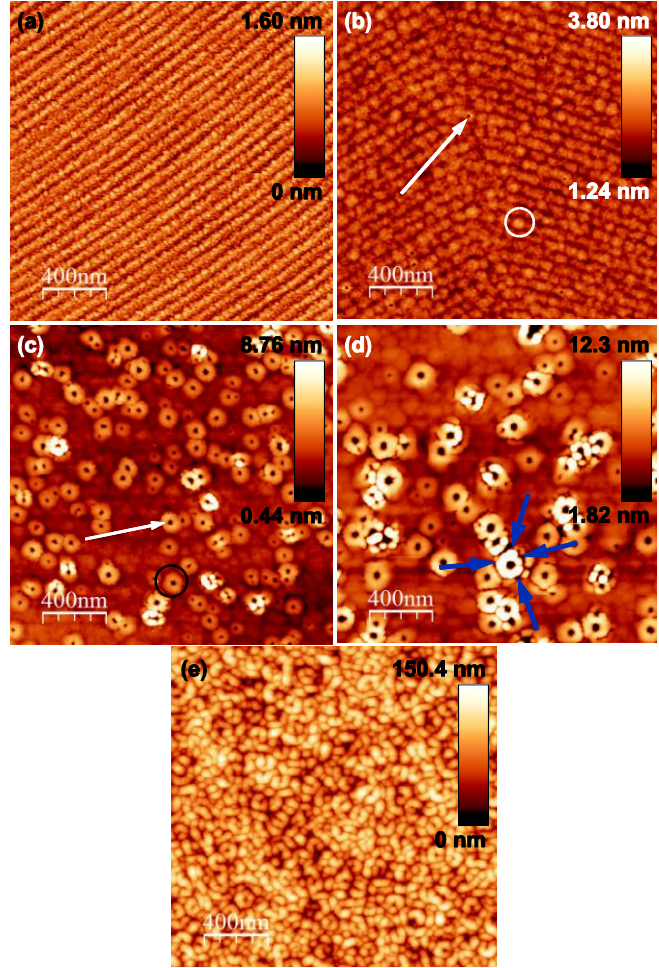

**Fig.1:** 2µm× 2µmAFM images of In$_x$Al$_{1-x}$N layers with an indium content of 19.7±0.8%grown on FS GaN substrates with an In/Al flux ratio of 3.5, for thicknesses of:(a) 2 nm,(b) 50 nm,(c) 100 nm, (d) 240 nm, and (e) 500 nm. Two V-defects are pointed by white arrows in (b) and (c). One hillock is circled in white in (b). One ring is circled in black in (c). Smaller V-defects surrounding rings are pointed by blue arrows in (d).

| Thickness (nm) | $x$ (%) | In/Al flux ratio | Hillock diameter (nm) | Hillock height (nm) | rms roughness (nm) | V-defect density (cm$^{-2}$) |
|---|---|---|---|---|---|---|
| 2 | - | 3.5 | - | - | 0.19 | 0 |
| 50 | 18.9 | 3.5 | 100 | 0.7 | 0.24 | 2.5 ×10$^8$ |
| 100 | 20.4 | 3.5 | 120 | 0.8 | 0.35 | 5 × 10$^9$ |
| 240 | 20.1 | 3.5 | 120 | 0.8 | 0.42 | 3 × 10$^9$ |
| 500 | 19.9 | 3.5 | - | - | 21.4 | - |
| 100 | 13.6 | 3.5 | 130 | 1 | 0.33 | 1× 10$^8$ |
| 210 | 13.5 | 3.5 | 110 (strong dispersion) | 1 | 0.38 | 2 × 10$^9$ |
| 50 | 19.5 | 1.8 | 90 | 1 | 0.31 | 5 × 10$^9$ |

**Table 1:** Main characteristics of In$_x$Al$_{1-x}$N layers, as deduced from AFM measurements, depending on the thickness, the indium composition, and the In/Al flux ratio.

Fig.1 shows AFM images of InAlN layers with an indium content of 19.7±0.8%, for thicknesses ranging from 2to 500 nm. For the 2 nm thick layer [Fig.1(a)], the atomic steps of the underlyingGaN substrateare still visible and the layer exhibits a rough morphology with a lot of small dots (10 to 20 nm in diameter), with a root mean square (rms) roughness of 0.19 nm. However, neither hillocks nor V-defects are observed. For the 50 nm thick layer [Fig.1(b)], hillocks with a diameter of about100 nm and a height of about 0.7 nmare aligned along atomic steps (one of these hillocks is circled in white), and therms roughness isincreased to 0.24 nm. A few V-defects appear with a low density of $2.5\times10^8$ cm$^{-2}$ (one V-defect is pointed by a white arrow). For the 100 nm thick layer [Fig.1(c)], the density of V-defects is increased to$5\times10^9$ cm$^{-2}$, which is more than two orders of magnitudelarger than the nominalthreading dislocation density.Slightly bigger hillocks with an average diameter of 120 nm and an average height of 0.8 nm are visible, and the rms roughness is increased to 0.35 nm (deduced from a measurement done for a small region without any V-defects). Moreover, ringslocated around the V-defects are also visible (one ring is circled in black), with an average height of about 3.5 nm and an average diameter of about 140 nm.For the 240 nm thick layer [Fig.1(d)], smaller V-defects pointed by blue arrows appear around the rings. These smaller V-defects can also be seen in the 100 nm thick sample, although being less visible. The density of V-defects is $3\times10^9$ cm$^{-2}$, which is approximately the same than for the 100 nm thick sample.The rms roughness is once more increased, to a value of 0.42 nm (value also deduced from a measurement done for a small region without any V-defects). Finally, for the 500 nm thick layer [Fig.1(e)], a very much rougher surface is observed with armsroughness of 21.4 nm. These numbers are compiled in Table 1. To summarize, hillocks appear from a thickness ranging between 2 and 50 nm and then V-defects progressively appear up toa thickness of 240 nm, finally leading to a very rough 500 nm thick layer.

## 2.Role of the growth temperature

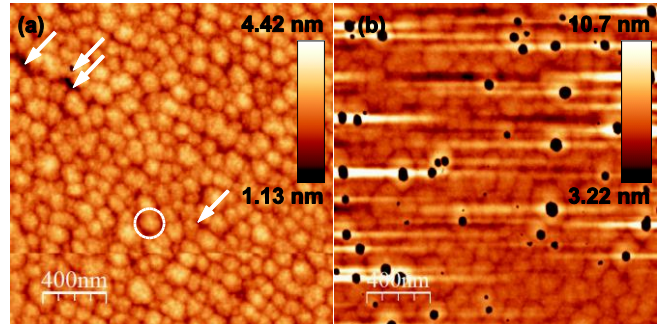

**Fig.2: 2μm× 2μmAFM images of InAlN layers with an indium content of 13.5% grown on FS GaN substrates with an In/Al flux ratio of 3.5, forthicknesses of (a) 100 nm, and (b) 210 nm. In (a), the four V-defects are indicated by white arrows, and one hillock is circled in white.**

Fig.2 shows AFM images of InAlN layers with 13.5% of In, for thicknesses of 100 nm [Fig.2(a)], and 210 nm [Fig.2(b)].For the 100 nm thick layer, hillocks are visible (one is circled in white) and the rms roughness amounts to 0.33 nm. Only fourV-defects (pointed by white arrows)are visible, which gives a V-defect density of $1\times 10^8$ cm$^{-2}$. For the 210 nm thick layer, the rms roughness is increased to 0.38 nm, and V-defects with a high density are visible ($2\times10^9$ cm$^{-2}$), but without any rings surrounding them, unlike for the 19.7% layers (as seen in Fig. 4 of Ref. 17: the InAlN layer under tensile strain does not exhibit rings contrary to the one under compressive strain).The critical thickness for the appearance of V-defects is thus higher for the 13.5% layers grown at higher temperature than for the 19.7% layers grown at lower temperature. The same tendency was also observed for InGaN layers grown on GaN: V-defects appear with increasing In content or thickness.[34] It was attributed to an increased strain and to a lower surface mobility, but the InAlN case shows that the same behavior is observed for nearly lattice-matched layers. For the 210 nm thick layer, cracks are observed on optical microscopy images (not shown), as it is usuallythe case for layers under tensile strain, and in particular for InAlN layers with an indium content lower than 17%.[17,35]

It is well known that in other III-V material systems, such as InGaAs grown on GaAs,compressive strain relaxation occurs by the formation of 3D islands (referred to as the Stranski-Krastanovgrowth mechanism).[36]Concerning the epitaxy of InGaAs layers on InP, it was reported that strain relaxation occurred in the case of tensile strainby the formation of valleys.[37]The rings around V-defects, which are visible for an indium content of 19.7% but not for 13.5%, could help to release the compressive strain. Because hillocks are present in the layers with an indium content of 19.7%, one could have thought that the hillock formationis due to relaxation of the compressive strain occurring by islanding.[38]However, as the same morphology is observed for layers under tensile strain with an indium content of 13.5%, another explanation to the hillock formation must be found. A similar morphology with such hillocks or mounds has already been observed for GaN layers grown at 800°C by ammonia-based molecular beam epitaxy.[39] The mounds were shown to disappear under annealing at 1000°C, giving flat terraces with an atomic step height. In our case, $In_{0.17}Al_{0.83}N$ is grown at 840°C, which is a compromise between In incorporation(larger at low temperature)and surface diffusion length of adatoms(larger at high temperature). The optimal growth temperature for AlN being 1100°C, we speculate that the mobility of adatoms is too low. Consequently, hillocks form because of kinetic roughening, *i.e.* a roughening due to an energy barrier present at low temperature, which prevents atoms to jump from the top of a hillock to its bottom.[39,40]

### 3. Role of the In/Al flux ratio

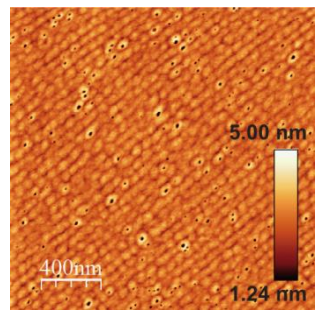

**Fig.3: 2μm× 2μm AFM image of a 50 nm thick InAlN layer with an indium content of 19.5% grown on a FS GaN substrate, with an In/Al flux ratio of 1.8.**

Another interesting feature is the role of the In/Al flux ratio on the morphology. We published in a previous work that a high In/Al flux ratio of 3.5 was necessary to make high quality DBRs, whereas with a low ratio of 1.8, interfaces were rough, and threading dislocations were created.[33] Fig.3 shows an AFM image of a 50 nm thick InAlN layer with an indium content of 19.5% grown with an In/Al ratio of 1.8. The hillocks are 1 nm high and 90 nm large, *i.e.* they have a smaller diameter and a larger height than the sample with the same thickness and with a flux ratio of 3.5 (equal to 0.7 nm and 100 nm, respectively), which is consistent with an increased kinetic roughening obtained for a lower In/Al flux ratio. The same dependency of the hillock diameter on the In/Al flux ratio was observed by another group.[41] The V-defect density is $5\times10^9$ cm$^{-2}$, 20 times larger than for the 50 nm thick layer grown with a ratio of 3.5. It is known that adsorbed species on a growing surface can modify the surface diffusion length, by the so-called surfactant effect.[42] Particularly, In was proposed to improve the Al surface mobility through this effect.[43] Thus increasing the In/Al flux ratio has the same rolethan increasing the temperature: it decreasesthe kinetic roughening,[41] and it delays the appearance of V-defects.Before going further in the discussion concerning the formation mechanism of V-defects, let us analyze their structure and the degradation mechanism of the layers with increasing thickness as deduced from TEM observations.

## B. Phase separation due to V-defects

The TEM observations have been performed on InAlN layers with an indium content of 19.7%± 0.8% for thicknesses of 100 and 500 nm. With a low density of threading dislocations in the FS GaN substrates (less than $4\times10^7$ cm$^{-2}$), the probability to observe them in a thin TEM lamella is small. Actually, no threading dislocations were observed by TEM in the studied samples.

### 1.V-defects appearance for 100 nm thick layers

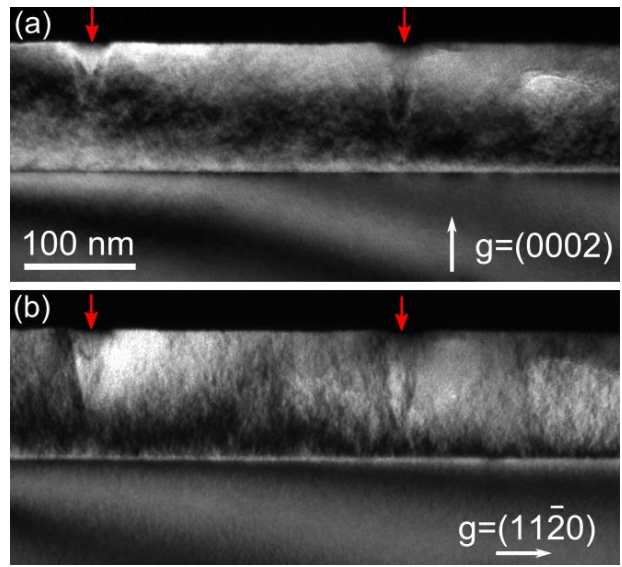

**Fig.4: Dark-field TEM cross-section images with (a) g=(0002) and (b) g=(11-20)of a 100 nm thick InAlN layer with an indium content of 19.7%, showing two V-defects, pointed by red arrows.**

Fig.4 showsdark-field TEM images taken with g=(0002) and g=(11-20) diffraction conditions of a 100 nm thick InAlN layer with an indium content of 19.7%. Two V-defects (indicated by red arrows) are visible without any structural defects such as threading dislocations, stacking mismatch boundaries, or stacking faults at their bottom. The same observation was done for all the V-defects we observed.This is consistent with the fact that the density of V-defects observed by AFM on this layer is two orders of magnitude larger than the nominal density of threading dislocations.Bothfor InGaN[34]and InAlN layers,[20–23] V-defects are generally attributed to threading dislocations. However, some authors have also mentioned the presence of V-defects not associated to any threading dislocations in InAlN layers.[17,23] Another interesting feature is the presence of sometriangular contrasts above the V-defects, as noted by Vennéguès *et al.*[23] Note that theslight grainy contrast visiblein Fig. 4 for the InAlN and GaN layers is due to the amorphous glue used during sample preparation that is covering the observed region. However, another contrast is visible in the InAlN layer and not inthe GaNsubstrate, especially when using theg=(11-20) diffraction condition. The same contrasts are visible in InAlN layers outside the V-defects on the plan-view images of Fig.5 andFig. 6. Sample preparation was done by wedge polishing using colloidal silica for the last polishing step, and ion milling was not used. The contrasts might be due to a selective attack by the colloidal silica of theInAlN layer. However, this contrast could also be due to some In composition fluctuations occurring at the scale of a few nanometers. In this latter case, it could be an explanation forthe large Stokes shift observed in InAlN layers.Indeed, InAlN lattice-matched to GaN was shown to absorb light at about 4.5 eV and to emit light under photoexcitation at 3.7 eV.[13,24] In this respect, it was pointed out in a recent theoretical paper that the numerous possible configurations for the In atoms within atom-supercells, each of them being characterized with different energy gaps,could lead to very different signatures in absorption or emission spectra.[44]

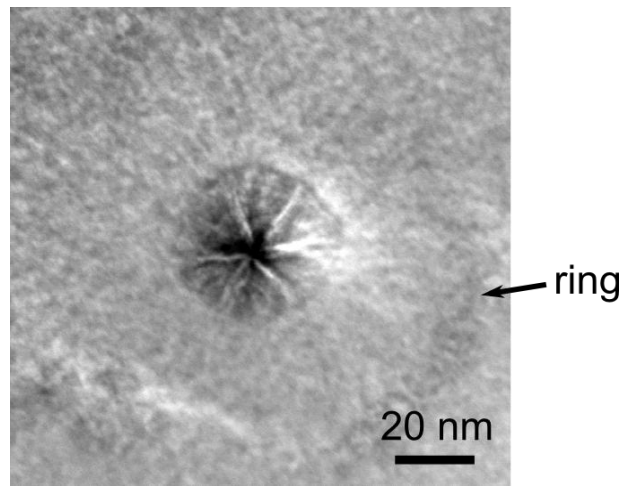

**Fig.5: Z-contrast STEM image of a plan-view100 nm thick InAlN layer with an indium content of 19.7%.**

Fig.5 is a Z-contrast STEM image in plan-view of the same sample.A V-defect with a diameter of about40 nm is visible. It exhibitssix bright branches, and a dark contrastinside, compared to the InAlN layer outside the V-defect.It is surrounded by a ring of about 100 nm in diameter, whose contrast is very weakbecause its thickness of 3 nm is relatively thin with respect to the thickness of the TEM lamella (estimated to rangebetween 10and 100 nm). Such V-defects present in InAlN layers havealready been deeply characterized. It was reported that the six facets of the V-defects were generally {11-2x} ones,[22,23,45,46] but for thick samples, {1-101} facets have also been observed.[23,47] It was shown that In preferentially segregates along the six<1-10x> branches of the V-defects, giving rise to In-rich vertical triangular portion of planes of the {11-20} type.[23] In the present case, the six facets are not very well defined, as can be observed in Fig.5. The same feature is observed for AFM images of the 19.7% layers (Fig.1(c) and 1(d)): V-defects seem circular, whereas for the 13.5% layers (Fig.2), V-defects seem well faceted. It could be due to the compressive strain, which leads to the formation of rings. We confirmed by EDX (not shown here) that the bright branches are indium-rich vertical planes, as described in the literature.[22,23,45,46] Concerning the dark contrast inside the V-defect, it could be due either to an In-poor region or to the fact that there is less matter inside the V-defect, which is an inverted empty pyramid.

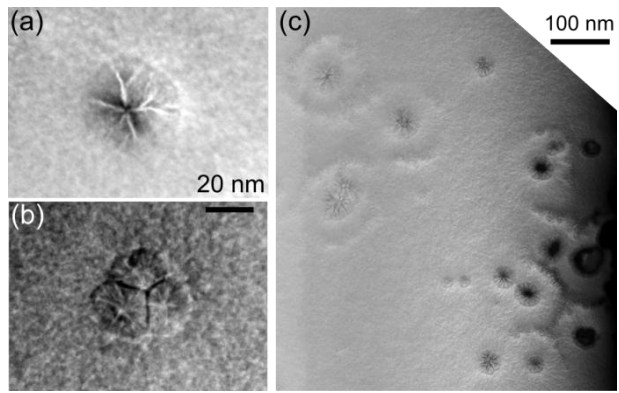

**Fig. 6: Plan-view of a 100 nm thick InAlN layer with an indium content of 19.7%. Z-contrast STEM images of (a) a V-defect different from the one shown in Fig.5, (b) three V-defects coalescing, and (c) big V-defects surrounded by smaller V-defects.**

Fig. 6 shows other plan-view images of the 100 nm thick InAlN sample with an indium content of 19.7%. Fig. 6(a) shows a V-defect with a different morphology from the one presented in Fig.5. There are initially five bright branches, two of them separating subsequently into two secondary branches. Thus it is likely that both <1-10x> and <11-2x> directions occur for these bright branches. Fig. 6(b) shows the coalescence of three V-defects. Three sets of bright In-rich lines are visible, together with three dark lines at the coalescence boundary between the V-defects. On these dark lines the lamella is not thinner than outside the V-defect, the dark contrast is consequently due to an In-poor InAlN region, which we attribute to a decrease in indium incorporation on the concave edges at the sides of the V-defects. We will develop this point in the next paragraph. Fig. 6(c) is a Z-contrast STEM image taken at lower magnification on a wedge-shaped TEM lamella, which is thinner on the right-hand side. It shows smaller V-defects around the rings, which surround bigger

V-defects. We have already highlighted this secondary nucleation of V-defects occurring around the first generationof V-defects in the AFM images (Fig.1(d)).

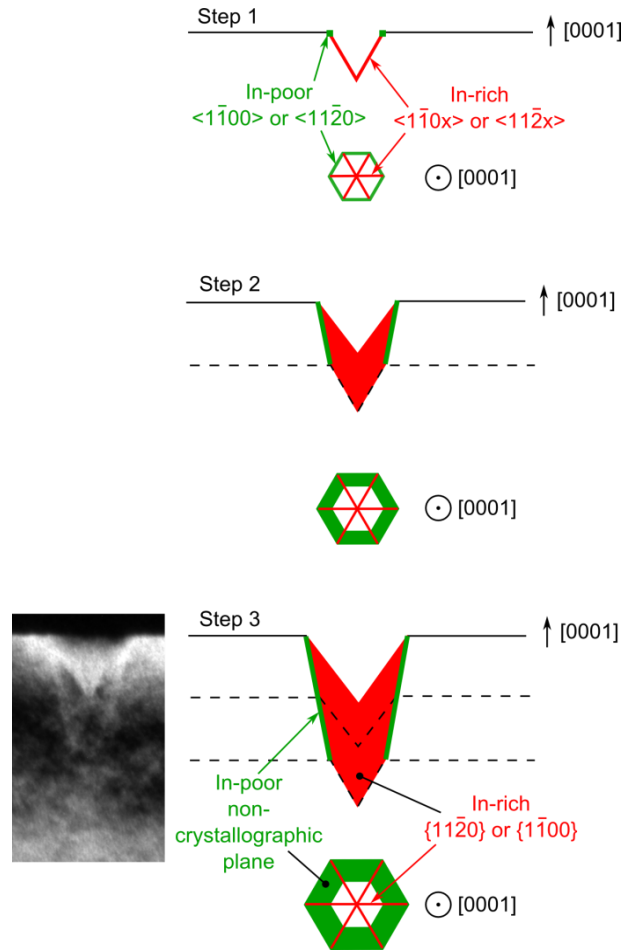

**Fig.7: 2D schematics of a V-defect displayed in cross-section and plan-view from its nucleation (step 1) to the final step (step 3). For step 3, a close-up view of the TEM image issued from Fig.4 is shown for comparison.**

Fig.7shows the growth steps of a V-defect with schematics displayed in cross-section and in plan-view. Inthe first step (step 1), a V-defect appears and preferential incorporation of In occurs on the concave parts inside the V-defects, *i.e.* on the <1-10x>or <11-2x>edges, drawn in red. At the same time, In incorporation is loweron the convex part outside the V-defect, *i.e.* on the <1-100> or <11-20> edges, drawn in green. It is worth noting that a similar observation was also pointed out forthe InAlNbarrier layers of GaN/InAlN multiple quantum wellswith the core-shell geometry grown onGaN nanocolumns oriented along the *c*-axis: on the 6 edges of the columns (which have a hexagonal base), In incorporation appears to be alsoweaker.[48]At this stage, let us remind that the optimal growth temperature forInNlayers by MOVPE is around 600°C, whereas it is around 1100°C for AlN. At the growth temperature considered here, *i.e.*around 840°C, Inincorporation is made difficult whereas it is not a problem for Al.We therefore propose that the desorption rate of In depends on the surface configuration: namely for concave surface shapes, In atoms would be more strongly linked, whereas for convex surface shapes, In atoms would desorb more easily, explaining the formation of In-rich and In-poor regions.To further validate this reasoning, we point out that for InAlN layers grown bymolecular beam epitaxya honeycomb structure with In-rich walls was observed and attributed to the preferential incorporation of In atoms between platelets formed at the beginning of the growth.[49]The driving force for such a process was attributed to the build-up of tensile strain between these coalescing platelets, but it could also be due to a better incorporation of In atoms because the surface between the platelets is concave. As the growth of the basal plane is faster than that of the V-defect inclined planes, the V-defect is growing during steps 2 and 3. The different incorporation ratesof In atoms leads to: (i) In-rich vertical triangular portion of planes of the {11-20} or {1-100} type drawn in red, and (ii) In-poor inclined planes drawn in green, which are not crystallographic planes but planes whose inclination is related to the ratio of the basal plane growth rate over the V-defect inclined plane growth rate. These In-rich and In-poor regions present at the bottom of the V-defects are visible in the TEM images shown inFig.4.

**2.Coalescence of V-defects from an InAlN layer thickness of 200 nm**

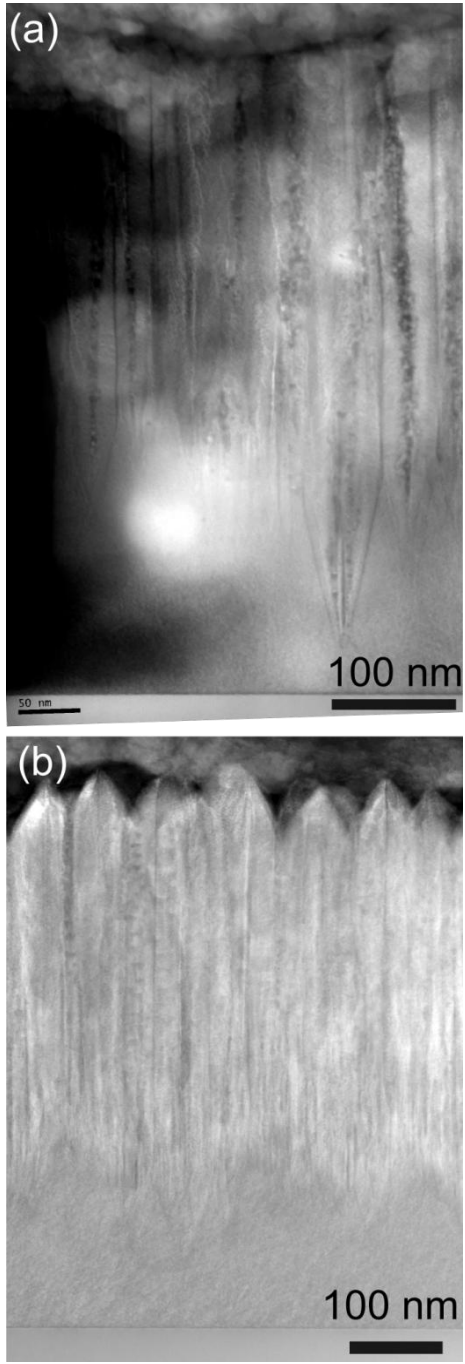

**Fig.8: Cross-section Z-contrast STEM images of a 500 nm thick InAlN layer with an indium content of 19.7%.**

Fig.8 shows cross-section Z-contrast STEM images for a 500 nm thick InAlN layer with 19.7% of In. The bottom InAlN layer is homogeneous, while the upper InAlN layer exhibits In-rich and In-poor regions, together with voids. The boundary between these two sub-layers is not abrupt because the upper layer is due to a growth occurring on coalesced V-defects, which will be described at the end of this section. At the surface, a *saw-tooth* morphology is observed, which corresponds to the very rough layer observed by AFM [Fig.1(e)]. This peculiar structure will be explained when describing the schematic shown in Fig.10. Once more, no threading dislocations coming from the FS GaN substrate are observed, which proves that all the coalesced V-defectswe haveobserved have another origin. From TEM imagestaken on a 3-μm wideregion, such as those shown in Fig. 8, it can be deduced that the very first V-defect appears for a 30 nm thickness and the very last for a 200 nm thickness, withthe majority of the V-defects appearing between 120 and 200 nm.

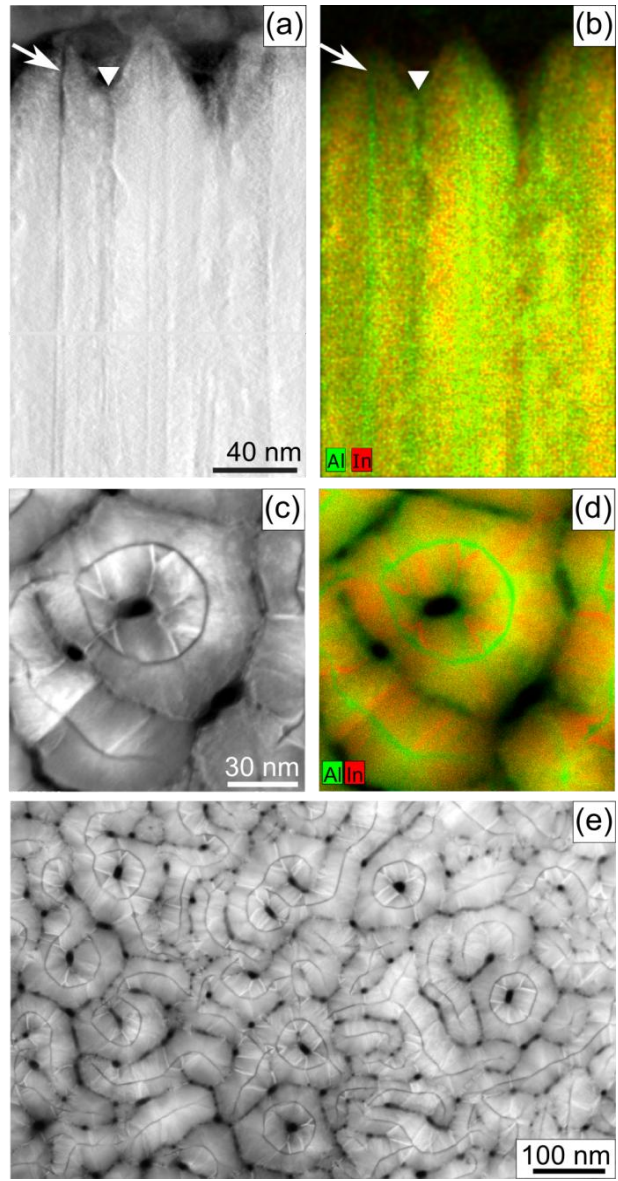

**Fig.9: 500 nm thick InAlN layer with an indium content of 19.7%. (a) Cross-section Z-contrast STEM image with (b) the associated EDX map. (c) Plan-view Z-contrast STEM image with (d) the associated EDX map.(e) Plan-view Z-contrast STEM image taken at a lower magnification.**

Fig.9shows Z-contrast STEM images of the500 nm thick InAlN layer with 19.7% of In, together with EDX maps.Fig.9(a) is a cross-section of the upper part of the InAlN layer, andFig.9(b) is the corresponding EDX map.On the Z-contrast image, there is always an ambiguity to determine whether a dark contrast is due to the presence of voids or to the presence of a lighter alloy, but EDX maps can solve this issue. Here a black vertical line indicated by an arrow is visible in a saw-tooth, and a larger black columnar contrast indicated by a triangle is visible in the pit between two saw-teeth. On the EDX map, the former contrast appears in green, it is thus In-poor. The latter contrast appears in dark, meaning that less matter is present, it is thus a void. Keeping this in mind, dark contrasts seen in Fig.8 can be ascribed to either a void or an In-poor region. The plan-view STEM image [Fig.9(c)] and the corresponding EDX map [Fig.9(d)] of the last 100 nm of the InAlN layer help to better understand the morphology of the structure. Voids are visible (appearing in black in the STEM image and in the EDX map), together with In-rich walls (bright lines in the STEM image, red lines in the EDX map) and In-poor walls (dark lines in the STEM image, green lines in the EDX map), as well as InAlN with a composition close to the nominal composition (grey background in the STEM image, yellow color in the EDX map). Fig.9(e) is a plan-view STEM image taken at a lower magnification. We have seen for the 100 nm thick InAlN sample that V-defects of the first generation were surrounded by a second generation of V-defects. Here, the first-generation of V-defects leads to bigger holes and are surrounded by a closed In-poor wall. The smaller V-defects lead to smaller holes, and they are boarded by In-poor walls too, but with a serpentine shape.

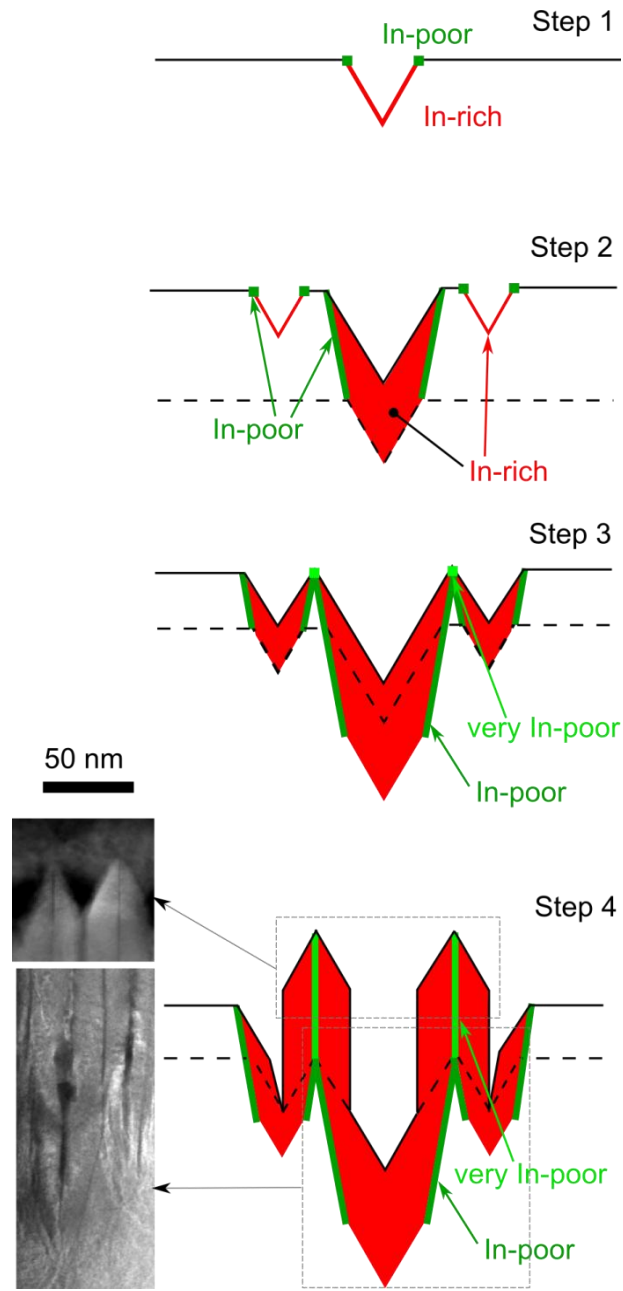

**Fig.10: Schematic cross-sections of coalescing V-defects. For step 4, Z-contrast STEM images of the 500 nm thick InAlN layer with an indium content of 19.7% are shown for the sake of illustration.**

Fig.10 shows the growth sequence of coalescing V-defects with schematic cross-sections. At the beginning (step 1), a V-defect nucleates and In-rich regions (shown in red) are formed at the concave edges, together with In-poor regions (shown in green) at the convex edges. Then growth proceeds further: the In-rich and In-poor regions expand while the V-defect is growing, and some smaller V-defects nucleate around the big one (step 2). At step 3, V-defects coalesce. We speculate that the In-poor region becomes even poorer because the angle between the surfaces is more acute. Then, this very In-poor region has a faster vertical growth rate, and growth of the In-rich regions and of InAlN close to the nominal composition regionsare pulled by this In-poor wall, leading to the formation of voids (step 4).This is visiblein the AFM imagedisplayed inFig.1(d):onceV-defects have coalesced, the surrounding rings are higher than when the V-defects are isolated. Z-contrast STEM images of the 500 nm thick InAlN layer with an indium content of 19.7% are also shownin Fig.10to further illustrate step 4.

## IV. DISCUSSION

We have mentioned in the introduction that a change inindium composition was observed during the growth of InAlN layers. It was previously found that a 65 nm thick InAlN layer[25] with 22% of In(but also for 60 nm[26],or 50 nm[27] thick InAlN layerswith 24% and 18% of In, respectively)was followed by an InAlN layerwith 26% of In (19% and 15% of In, respectively). It was noted that this change in composition occursonly for the lowest growth temperatures.One group proposed a mechanism for this degradation, which is driven by threading dislocations.[29,30] When growth is done on a GaN template grown on a foreign substrate, threading dislocations are indeed present in the GaN template with a high density ($10^8$ to $10^9$ cm$^{-2}$). When growing InAlN on such atemplate, V-defects appear because of the presence of threading dislocations. With further growth, V-defects coalesce and because of growth occurring on the inclined facets of the V-defects, an In-poor InAlN layer is obtained.[29,30] However, in our case for growthsperformed on high-quality FS GaN substrates, the V-defects observed by TEM are not correlated with the presence of threading dislocations. Concerning the change in composition occurring because of the growth on top ofcoalesced V-defects, we can be more precise by saying that In-rich and In-poor walls are present on top of the V-defects, because of the different indium incorporation on the concave and convex parts of the V-defects. Depending on the size of the V-defects, it could lead to either a globally richer[25] or poorer[26,27] degraded top layer.

Let us discuss now what are the possible nucleation mechanisms for the V-defects. These defects have been observed in GaN, InGaN, and InAlN layers. Most often, threading dislocations (especially those with a Burgers vector with a screw component) are invoked to be at the origin of V-defects for InGaN[34] or InAlN layers.[20–23] Other possibilities for the nucleationof V-defects have been proposed. Dopants were shown to increase the V-defect density in GaN[50] and InAlN layers.[10] In InGaN, a stacking mismatch boundary occurring on top of a stacking fault was correlated with the presence of a V-defect.[51] Finally, V-defects due to dislocations were shown to release the compressive strain in thick InGaN layers, by increasing their size.[52,53] In our case, V-defects are not due to threading dislocations or to another structural defect such as a stacking mismatch boundary or a stacking fault. Because they are visible for layers both under tensile and compressive strain, they are likely not due to strain. Non intentional impurities such as carbon or oxygen may play a role,[19] but it cannot explain the occurrence of V-defects from a certain critical thickness, which depends on the growth temperature and the In flux. Actually, we have shown that V-defects appear for thinner layers when the surface diffusion length of species is shorter, *i.e.* at lower growth temperatures or with a low In/Al flux ratio (without surfactant effect improving the adatomdiffusion). Instead we propose that hillocks are at the origin of the V-defects. We remind that hillocks are observed for both layers under tensile and compressive strain, and we attributed their formation to kinetic roughening: with an increased surface diffusion length, hillocks are thinner and larger. When the layer becomes thicker, the diameter and height of hillocks increase. There would be a critical height forthose hillocksbeyond which inclined facets between hillocks become stable, giving rise to V-defects. This would explain theirprogressive nucleation on the whole sample surface, occurring for a thickness ranging between 120 and 200 nm for InAlN layers with 19.7% of In. Moreover, the valleys around rings are deep (around 1 nm), and they will consequently act as preferential nucleation sites for V-defects. The apparition of V-defects is delayed for higher temperatures or when the surfactant effect is at play, which can be explained then by an enhanced diffusion, which givesrise to smaller hillocks.

To circumvent the formation of V-defects and the subsequent degradation of InAlN layers, the adatomdiffusion length should be increased. One key parameter to play with is the In/Al flux ratio,[33] which increases the surface diffusion length via the surfactant effect. Increasing the temperature is another solution, but a too large increase lead to cracked layers because of a lower indium content. Always with the aimof improving diffusion, amore efficient surfactant than In could be used.[42] To prevent the formation of hillocks, growth on substrates with a large miscut could also be tested in order to have shorter terraces than the hillock diameter.[40,41] Growth on polar or semi-polar planes should not produce V-defects, which are defects with a hexagonal symmetry specific to *c*-plane growth. However, in this latter case it is likely that kinetic roughening leading to other defects would also occur. Finally, we point out that multilayer structures with GaN interlayers grown at high temperature and InAlN layers thinner than the critical thickness for theappearance of V-defects should lead toa good structural quality. Depending on the targeted application, this could be an attractive solution.

## V. SUMMARY AND CONCLUSIONS

In this paper, we have discussed the growth mechanisms of thick nearly lattice-matched InAlN layers grown on FS GaN substrates, either under a slight compressive or tensile strain. First, we discussed the origin of the hillocks (typically characterized by a 100 nm diameter and a 1 nm height), observed for layers under tensile and compressive strain. We attribute these hillocks to kinetic roughening, *i.e.* to an energy barrier preventing atoms from jumping down anatomic step. Second, we observed V-defects whatever the type of strain (compressive or tensile), and theyarein our case not due to threading dislocations(growthsperformed on high-quality FS GaN substrates). They appear for a certain critical thickness, which increases when increasing the growth temperature or the In/Al flux ratio, *i.e.* when increasing the surface diffusion length. We proposed that they are due to stabilized inclined facets for a critical size of the hillocks. Third, it is shown that phase separation occurs because of a better (worse) incorporation of In taking place at the concave (convex) parts of the V-defects leading to the formation of In-rich (poor) regions. When V-defects coalesce, In-rich and In-poor walls, InAlN with a composition close to the nominal composition and columnar voids are formed, giving rise to a rough phase-separated upper layer on the initially homogeneous InAlN layer.

This work provides useful insights on thedegradation mechanism of thick nearly lattice-matched InAlN layers, but it can also be useful to understand the degradation mechanism of thick strained InGaN layers. Indeed, thick InGaN layers present a similar morphology to that of thick InAlN layers.[31,54] The degradationobservedwith increasing thickness is generally attributed to strain relaxation,[31,34] and/or to V-defects due to threading dislocations,[34] and/or to phase separation due to spinodal decomposition.[55] The presentwork points out that even in the absence of stress and threading dislocations, In-rich nitride alloys grown with the (0001) orientation have a strong tendency to form V-defects, which can eventually lead to phase separation.This phase separation is not due to spinodal decomposition ($In_{0.17}Al_{0.83}N$ was shown to be stable up to 960°C)[56] but to a different incorporation of indium on concave or convex shape surfaces of the V-defects.

## ACKNOWLEDGMENTS

The authors thank M. Cantoni, D. Alexander, and C. Hébert from the CIME laboratory at EPFL for opening access to the TEM facilities and for their help during TEM experiments.This work was supported by the NCCR Quantum Photonics (NCCR QP), research instrument of the Swiss National Science Foundation (SNSF), and by Rainbow (Contract No. PITN-GA-2008-213238), a Marie Curie initial training network under the 7$^{th}$ framework program, funded by the European Commission.

**REFERENCES**

[1] J.-F. Carlin and M. Ilegems, Appl. Phys. Lett. **83**, 668 (2003).

[2] J. Kuzmik, IEEE Electron. Device Lett. **22**, 510 (2001).

[3] A. Dadgar, F. Schulze, J. Bläsing, A. Diez, A. Krost, M. Neuburger, E. Kohn, I. Daumiller, and M. Kunze, Appl. Phys. Lett. **85**, 5400 (2004).

[4] M. Gonschorek, J.-F. Carlin, E. Feltin, M. A. Py, and N. Grandjean, Appl. Phys. Lett. **89**, 062106 (2006).

[5] H. Sun, A.R. Alt, H. Benedickter, E. Feltin, J.-F. Carlin, M. Gonschorek, N. Grandjean, and C.R. Bolognesi, IEEE Electron. Device Lett. **31**, 957 (2010).

[6] S. Nicolay, J.-F. Carlin, E. Feltin, R. Butté, M. Mosca, N. Grandjean, M. Ilegems, M. Tchernycheva, L. Nevou, and F. H. Julien, Appl. Phys. Lett. **87**, 111106 (2005).

[7] S. Senda, H. Jiang, and T. Egawa, Appl. Phys. Lett. **92**, 203507 (2008).

[8] A. Castiglia, J.-F. Carlin, E. Feltin, G. Cosendey, J. Dorsaz, and N. Grandjean, Appl. Phys. Lett. **97**, 111104 (2010).

[9] R. Charash, H. Kim-Chauveau, J.-M. Lamy, M. Akther, P. P. Maaskant, E. Frayssinet, P. de Mierry, A. D. Dräger, J.-Y. Duboz, A. Hangleiter, and B. Corbett, Appl. Phys. Lett. **98**, 201112 (2011).

[10] Y. Taniyasu, J.-F. Carlin, A. Castiglia, R. Butté, and N. Grandjean, Appl. Phys. Lett. **101**, 082113 (2012).

[11] X. Zeng, D. L. Boïko, G. Cosendey, M. Glauser, J.-F. Carlin, and N. Grandjean, Appl. Phys. Lett. **101**, 141120 (2012).

[12] G. Cosendey, A. Castiglia, G. Rossbach, J.-F. Carlin, and N. Grandjean, Appl. Phys. Lett. **101**, 151113 (2012).

[13] R. Butté, J.-F. Carlin, E. Feltin, M. Gonschorek, S. Nicolay, G. Christmann, D. Simeonov, A. Castiglia, J. Dorsaz, H. J. Buehlmann, S. Christopoulos, G. Baldassarri Höger von Högersthal, A. J. D. Grundy, M. Mosca, C. Pinquier, M. A. Py, F. Demangeot, J. Frandon, P. G. Lagoudakis, J. J. Baumberg, and N. Grandjean, J. Phys. D: Appl. Phys. **40**, 6328 (2007).

[14] *III-nitride Semiconductors and Their Modern Devices*, edited by B. Gil, (Oxford University Press, Oxford, 2013).

[15] K. Lorenz, N. Franco, E. Alves, S. Pereira, I.M. Watson, R.W. Martin, and K.P. O'Donnell, J. Cryst. Growth **310**, 4058 (2008).

[16] T.C. Sadler, M.J. Kappers, and R.A. Oliver, J. Cryst. Growth **311**, 3380 (2009).

[17] Z. L. Miao, T. J. Yu, F. J. Xu, J. Song, L. Lu, C. C. Huang, Z. J. Yang, X. Q. Wang, G. Y. Zhang, X. P. Zhang, D. P. Yu, and B. Shen, J. Appl. Phys. **107**, 043515 (2010).

[18] H. Kim-Chauveau, P. de Mierry, J.-M. Chauveau, and J.-Y. Duboz, J. Cryst. Growth **316**, 30 (2011).

[19] R. B. Chung, F. Wu, R. Shivaraman, S. Keller, S. P. DenBaars, J. S. Speck, and S. Nakamura, J. Cryst. Growth **324**, 163 (2011).

[20] Z. L. Miao, T. J. Yu, F. J. Xu, J. Song, C. C. Huang, X. Q. Wang, Z. J. Yang, G. Y. Zhang, X. P. Zhang, D. P. Yu, and B. Shen, Appl. Phys. Lett. **95**, 231909 (2009).

[21] J. Song, F. J. Xu, X. D. Yan, F. Lin, C. C. Huang, L. P. You, T. J. Yu, X. Q. Wang, B. Shen, K. Wei, and X. Y. Liu, Appl. Phys. Lett. **97**, 232106 (2010).

[22] A. Mouti, J.-L. Rouvière, M. Cantoni, J.-F. Carlin, E. Feltin, N. Grandjean, and P. Stadelmann, Phys. Rev. B **83**, 195309 (2011).

[23] P. Vennéguès, B.S. Diaby, H. Kim-Chauveau, L. Bodiou, H.P.D. Schenk, E. Frayssinet, R.W. Martin, and I.M. Watson, J. Cryst. Growth **353**, 108 (2012).

[24] J. -F. Carlin, C. Zellweger, J. Dorsaz, S. Nicolay, G. Christmann, E. Feltin, R. Butté, and N. Grandjean, Phys. Stat. Sol. (b) **242**, 2326 (2005).

[25] V. Darakchieva, M. Beckers, M.-Y. Xie, L. Hultman, B. Monemar, J.-F. Carlin, E. Feltin, M. Gonschorek, and N. Grandjean, J. Appl. Phys. **103**, 103513 (2008).

[26] K. Lorenz, S. Magalhães, N. Franco, N. P. Barradas, V. Darakchieva, E. Alves, S. Pereira, M. R. Correia, F. Munnik, R. W. Martin, K. P. O'Donnell, and I. M. Watson, Phys. Stat. Sol. (b) **247**, 1740 (2010).

[27] A. Redondo-Cubero, K. Lorenz, R. Gago, N. Franco, M.-A. di Forte Poisson, E. Alves, and E. Muñoz, J. Phys. D: Appl. Phys. **43**, 055406 (2010).

[28] S. Kret, A. Wolska, M. T. Klepka, A. Letrouit, F. Ivaldi, A. Szczepańska, J.-F. Carlin, N. A. K. Kaufmann, and N. Grandjean, J. Phys.: Conf. Ser. **326**, 012013 (2011).

[29] Q. Y. Wei, T. Li, Y. Huang, J. Y. Huang, Z. T. Chen, T. Egawa, and F. A. Ponce, Appl. Phys. Lett. **100**, 092101 (2012).

[30] Z. T. Chen, K. Fujita, J. Ichikawa, and T. Egawa, J. Appl. Phys. **111**, 053535 (2012).

[31] K. Pantzas, G. Patriarche, G. Orsal, S. Gautier, T. Moudakir, M. Abid, V. Gorge, Z. Djebbour, P. L Voss, and A. Ougazzaden, Phys. Stat. Sol. (a) **209**, 25 (2012).

[32] K. Lorenz, N. Franco, E. Alves, I. M. Watson, R. W. Martin, and K. P. O'Donnell, Phys. Rev. Lett. **97**, 085501 (2006).

[33] G. Cosendey, J-F. Carlin, N. A. K. Kaufmann, R. Butté, and N. Grandjean, Appl. Phys. Lett. **98**, 181111 (2011).

[34] I.-H. Kim, H.-S. Park, Y.-J. Park, and T. Kim, Appl. Phys. Lett. **73**, 1634 (1998).

[35] T. C. Sadler, M. J. Kappers, and R. A. Oliver, J. Cryst. Growth **314**, 13 (2011).

[36] D. Leonard, M. Krishnamurthy, C. M. Reaves, S. P. Denbaars, and P. M. Petroff, Appl. Phys. Lett. **63**, 3203 (1993).

[37] P. Krapf, Y. Robach, M. Gendry, and L. Porte, J. Cryst. Growth **181**, 337 (1997).

[38] A. Krost, C. Berger, P. Moser, J. Bläsing, A. Dadgar, C. Hums, T. Hempel, B. Bastek, P. Veit, and J. Christen, Semicond. Sci. Technol. **26**, 014041 (2011).

[39] S. Vézian, F. Natali, F. Semond, and J. Massies, Phys. Rev. B **69**, 125329 (2004).

[40] M. D. Johnson, C. Orme, A. W. Hunt, D. Graff, J. Sudijono, L. M. Sander, and B. G. Orr, Phys. Rev. Lett. **72**, 116 (1994).

[41] D. F. Brown, S. Keller, T. E. Mates, J. S. Speck, S. P. DenBaars, and U. K. Mishra, J. Appl. Phys. **107**, 033509 (2010).

[42] J. Massies and N. Grandjean, Phys. Rev. B **48**, 8502 (1993).

[43] S. Keller, S. Heikman, I. Ben-Yaacov, L. Shen, S. P. DenBaars, and U. K. Mishra, Appl. Phys. Lett. **79**, 3449 (2001).

[44] L. C. de Carvalho, A. Schleife, J. Furthmüller, and F. Bechstedt, Phys. Rev. B **85**, 115121 (2012).

[45] Th. Kehagias, G. P Dimitrakopulos, J. Kioseoglou, H. Kirmse, C. Giesen, M. Heuken, A. Georgakilas, W. Neumann, Th. Karakostas, and Ph. Komninou, Appl. Phys. Lett. **95**, 071905 (2009).

[46] Th. Kehagias, G. P. Dimitrakopulos, J. Kioseoglou, H. Kirmse, C. Giesen, M. Heuken, A. Georgakilas, W. Neumann, Th. Karakostas, and Ph. Komninou, Appl. Phys. Lett. **99**, 059902 (2011).

[47] L. Zhou, M. R. McCartney, D. J. Smith, A. Mouti, E. Feltin, J.-F. Carlin, and N. Grandjean, Appl. Phys. Lett. **97**, 161902 (2010).

[48] C. Bougerol, Private communication (2012).

[49] S. -L. Sahonta, G. P. Dimitrakopulos, Th. Kehagias, J. Kioseoglou, A. Adikimenakis, E. Iliopoulos, A. Georgakilas, H. Kirmse, W. Neumann, and Ph. Komninou, Appl. Phys. Lett. **95**, 021913 (2009).

[50] Z. Liliental-Weber, Y. Chen, S. Ruvimov, and J. Washburn, Phys. Rev. Lett. **79**, 2835 (1997).

[51] H. K. Cho, J. Y. Lee, G. M. Dalpian, and C. S. Kim, Appl. Phys. Lett. **79**, 215 (2001).

[52] T. L. Song, J. Appl. Phys. **98**, 084906 (2005).

[53] D. Won, X. Weng, and J. M. Redwing, J. Appl. Phys. **108**, 093511 (2010).

[54] K. Pantzas, G. Patriarche, D. Troadec, S. Gautier, T. Moudakir, S. Suresh, L. Largeau, O. Mauguin, P. L. Voss, and A. Ougazzaden, Nanotechnology **23**, 455707 (2012).

[55] F. A. Ponce, S. Srinivasan, A. Bell, L. Geng, R. Liu, M. Stevens, J. Cai, H. Omiya, H. Marui, and S. Tanaka, Phys. Stat. Sol. (b) **240**, 273 (2003).

[56] A. Gadanecz, J. Bläsing, A. Dadgar, C. Hums, and A. Krost, Appl. Phys. Lett. **90**, 221906 (2007).